\documentclass[twocolumn, prl, reprint, nofootinbib]{revtex4-1}

\usepackage{amsmath, amsfonts, amssymb}
\usepackage{natbib}
\usepackage{mathrsfs}
\usepackage[hidelinks]{hyperref}

\hypersetup{
  colorlinks   = true, 
  urlcolor     = blue, 
  linkcolor    = blue, 
  citecolor   = blue 
}

\begin{document}

\title{Construction of all $N\!=\!4$ conformal supergravities}

\author{Daniel Butter$^{1}$, Franz Ciceri$^{1}$, Bernard de
  Wit$^{1,2}$ and Bindusar Sahoo$^{3}$}
\affiliation{$^{1}$ Nikhef, Science Park 105, 1098 XG Amsterdam, The
  Netherlands\\
$^{2}$ Institute for Theoretical Physics, Utrecht University,
Princetonlaan 5, 3584 CC Utrecht, The Netherlands \\ 
$^{3}$ Indian Institute of Science Education and Research,
Thiruvananthapuram, Kerala 695016, India} 

\begin{abstract}
  All $N\!=\!4$ conformal supergravities in four space-time dimensions are
  constructed. These are the only $N\!=\!4$ supergravity theories whose
  actions are invariant under off-shell supersymmetry. They are
  encoded in terms of a holomorphic function that is homogeneous of
  zeroth degree in scalar fields that parametrize an
  $\mathrm{SU}(1,1)/\mathrm{U}(1)$ coset space. When this function
  equals a constant the Lagrangian is invariant under continuous
  $\mathrm{SU}(1,1)$ transformations. The construction of these
  higher-derivative invariants also opens the door to various applications
  for non-conformal theories.
\end{abstract}

\maketitle
\allowdisplaybreaks

Conformal supergravity is the supersymmetric generalization of
conformal gravity, whose Lagrangian is the square of the Weyl
tensor. The combination of local supersymmetry and conformal symmetry necessarily implies the presence of additional local invariances,
which include a special supersymmetry known as S-supersymmetry. The
original supersymmetry is then called Q-supersymmetry. These
combined invariances are associated with a superconformal gauge
algebra known as $\mathfrak{su}(2,2\vert N)$.  The transformation
rules and the corresponding invariant Lagrangians are known for
$N\!=\!1$ and $2$ \cite{Kaku:1978nz, deWit:1979dzm}. For $N\!=\!4$ the
full non-linear transformation rules of the fields, which constitute
the so-called Weyl supermultiplet, have been determined
\cite{Bergshoeff:1980is}. This is the largest possible conformal
supergravity that can exist in four space-time dimensions
\cite{deWit:1978pd}, and so far a complete Lagrangian was not known. A
unique feature is the presence of dimensionless scalar fields that
parametrize an $\mathrm{SU}(1,1)/\mathrm{U}(1)$ coset space. The
$\mathrm{U}(1)$ factor is realized as a local symmetry with a
composite connection, which acts chirally on the fermions. Hence the
so-called R-symmetry group is extended to
$\mathrm{SU}(4)\times\mathrm{U}(1)$.

As explained below there are good reasons to expect that a large
variety of these theories will exist. This Letter reports important
progress on this question as we derive the most general invariant
Lagrangian, which turns out to depend on a single arbitrary holomorphic
and homogeneous function of the coset fields. Here we will present its
purely bosonic terms; full results will be reported elsewhere. When
this function is constant these bosonic terms turn out to agree with a
recent result derived by imposing supersymmetry on terms that are at
most quadratic in the fermions \cite{Ciceri:2015qpa}.

In the Lagrangian this function will, for instance, multiply the terms
quadratic in the Weyl tensor. The possible existence of such a
non-minimal coupling was suggested long ago in \cite{Fradkin:1981jc,
  Fradkin:1982xc}. Meanwhile indirect evidence came from string
theory, where the threshold corrections in the effective action of IIA
string compactifications on $\mathrm{K3}\times T^2$ reveal the
presence of terms proportional to the square of the Weyl tensor
multiplied by a modular function \cite{Harvey:1996ir}. The same terms
emerge in the semiclassical approximation of microscopic degeneracy
formulae for dyonic BPS black holes
\cite{Dijkgraaf:1996it,LopesCardoso:2004law,Jatkar:2005bh}. Finally
higher-derivative couplings derived for $N\!=\!4$ Poincar\'e
supergravity \cite{Bossard:2012xs,Bossard:2013rza} do also exhibit
non-trivial scalar interactions. The results of this Letter can
provide more detailed information on such higher-derivative
interactions. Likewise they can be utilized to study the subleading
contributions to $N\!=\!4$ BPS black hole entropy in a fully
supersymmetric description.
 
We briefly summarize the field content of $N\!=\!4$ conformal
supergravity, which comprises $128+128$ bosonic and
fermionic off-shell degrees of freedom. Space-time indices are denoted by
$\mu,\nu,\ldots$, tangent space indices by $a,b,\ldots$ and
$\mathrm{SU}(4)$ indices by $i,j,\ldots$.  Among the bosonic fields
are the vierbein $e_\mu{\!}^a$, $\mathrm{SU}(4)$ gauge fields
$V_\mu{\!}^i{}_j$ and a gauge field $b_\mu$ associated with
dilatations. Furthermore there are three composite bosonic gauge
fields, namely the spin connection $\omega_\mu{\!}^{ab}$, the gauge
field $f_\mu{\!}^a$ associated with conformal boosts and the gauge
field $a_\mu$ associated with the $\mathrm{U}(1)$ symmetry.  In
addition to the gauge fields, the bosonic fields comprise complex
anti-selfdual tensor fields $T_{ab}{\!}^{ij}$ transforming in the
$\bf6$ representation of $\mathrm{SU}(4)$, whose complex conjugates
are the selfdual fields $T_{ab ij}$, complex scalars $E_{ij}$, and
pseudo-real scalars $D^{ij}{\!}_{kl}$ transforming in the
$\overline{\bf 10}$ and the ${\bf 20}'$ representation of
$\mathrm{SU}(4)$, respectively. Finally there exists a doublet of
complex scalars $\phi^\alpha$, which are invariant under dilatations
and transform under rigid $\mathrm{SU}(1,1)$ transformations ($\alpha=
1,2$). They are subject to the $\mathrm{SU}(1,1)$ invariant
constraint,
\begin{equation}
\phi^\alpha\,\phi_\alpha  = 1\,,\quad 
\phi_1\equiv (\phi^1)^\ast\,,\quad  \phi_2\equiv -(\phi^2)^\ast\,,
\end{equation}
and as a result the fields $\phi^\alpha$ and $\phi_\alpha$ parametrize
$\mathrm{SU}(1,1)$ matrices. Because these fields are subject to the
local $\mathrm{U}(1)$ symmetry, they describe two physical degrees of freedom
associated with an $\mathrm{SU}(1,1)/\mathrm{U}(1)$ coset space. 

The positive chirality fermions consist of the gravitini
$\psi_\mu{\!}^i$ (the gauge fields of Q-supersymmetry), a composite
gauge field $\phi_{\mu{i}}$ associated with S-supersymmetry,
and two spinor fields, $\Lambda_i$ and $\chi^{ij}{\!}_k$, transforming
in the $\bf4$, the $\overline{\bf4}$, the $\overline{\bf4}$ and the
$\bf{20}$ representation, respectively. The negative chirality
fermions transform under the corresponding conjugate representations.

We will refrain from discussing the superconformal transformations of
the various fields in any detail. These results have already appeared
in \cite{Bergshoeff:1980is,Ciceri:2015qpa}. The Weyl multiplet defines
a doublet of chiral superfields with lowest components $\phi_\alpha$,  
but no other independent chiral supermultiplets can exist coupled to
conformal supergravity.

In view of their relevance for the present paper we first present some
further details regarding the fields $\phi_\alpha$ and $\phi^\alpha$,
which we refer to as the holomorphic and the anti-holomorphic fields,
respectively. The holomorphic fields carry $\mathrm{U}(1)$ charge
equal to $-1$ and transform under Q-supersymmetry into the positive
chirality spinors $\Lambda_i$, which themselves carry $\mathrm{U}(1)$
charge $-3/2$,
\begin{equation}
  \label{eq:phi-transf}
  \phi_\alpha \to \mathrm{e}^{-\mathrm{i} \lambda(x)}
  \,\phi_\alpha\,,\qquad \delta\phi_\alpha= -\bar\epsilon^i
  \Lambda_i\, \varepsilon_{\alpha\beta} \,\phi^\beta\,.
\end{equation}
The supercovariant constraint that determines the $\mathrm{U}(1)$
gauge field $a_\mu$ and the generalized supercovariant derivatives of
the coset fields, $P_a$ and $\bar P_a$, are defined by
\begin{align}
\label{eq:U(1)-gauge}
    \phi^\alpha
    D_a\phi_\alpha=&\;-\tfrac14\bar\Lambda^i\gamma_a\Lambda_i\,, \\
        P_a=&\; \phi^\alpha\,\varepsilon_{\alpha\beta}
    \,D_a\phi^\beta\,,  \quad 
    \bar P_a= -\phi_\alpha\,\varepsilon^{\alpha\beta}
    \,D_a\phi_\beta\,, \nonumber
\end{align}
where $D_a$ denotes the fully superconformal covariant derivative.
Note that $P_a$ and $\bar P_a$ carry Weyl weight $+1$ and
$\mathrm{U}(1)$ weights $+2$ and $-2$, respectively. 
From these definitions one may derive the supercovariant
extension of the Maurer-Cartan equations associated with the
$\mathrm{SU}(1,1)/\mathrm{U}(1)$ coset space,
\begin{align}
  \label{eq:super-MC}
  F(a)_{ab} =&\, -2\mathrm{i}\, P_{[a}\,\bar P_{b]}
  - \tfrac12\mathrm{i}
  \big(\bar\Lambda^i\gamma_{[a}D_{b]}\Lambda_i-\text{h.c.}\big) \,,
  \nonumber \\
  D_{[a} P_{b]}=&\, -\tfrac12\bar\Lambda_i\gamma_{[a}\Lambda^i \,
  P_{b]}   +\tfrac14\bar\Lambda^i\,R(Q)_{ab\,i} \,,
\end{align}
where $F(a)_{ab}$ and $R(Q)_{ab\,i}$ denote the supercovariant
$\mathrm{U}(1)$ and Q-supersymmetry curvatures, respectively.

Note that the expressions \eqref{eq:phi-transf} and
\eqref{eq:U(1)-gauge}, when combined with those for the
anti-holomorphic fields, reflect the structure of the three left-invariant vector
fields associated with the group $\mathrm{SU}(1,1)$,
\begin{align}
  \label{eq:left-invariant-su11}
  \mathcal{D}^0=&\; \phi^\alpha\,
  \frac{\partial}{\partial\phi^\alpha}- \phi_\alpha\,
  \frac{\partial}{\partial\phi_\alpha} \,,\\
    \mathcal{D}^\dagger=&\; \phi_\alpha\,\varepsilon^{\alpha\beta} \,
  \frac{\partial}{\partial\phi^\beta}\;,\qquad 
   \mathcal{D}= -\phi^\alpha\,\varepsilon_{\alpha\beta} \,
  \frac{\partial}{\partial\phi_\beta}\;, \nonumber
\end{align}
which satisfy the commutation relations $\big[\mathcal{D}^0  ,
\mathcal{D}\big] = 2 \, \mathcal{D}$ and
$\big[\mathcal{D}, \mathcal{D}^\dagger\big] =  \mathcal{D}^0$. Using
these definitions the supersymmetry variation
and the supercovariant derivative of arbitrary functions 
$\mathcal{H}(\phi_\alpha,\phi^\beta)$ can be written as
\begin{align}
  \label{eq:der-H}
  \delta\mathcal{H} =&\; -\big[\bar\epsilon^i\Lambda_i\,\mathcal{D} +
  \bar\epsilon_i\Lambda^i\,\mathcal{D}^\dagger \big]
  \,\mathcal{H}\,,\nonumber\\
    D_a\mathcal{H} =&\; \big[ \bar P_a\,\mathcal{D} +
   P_a\,\mathcal{D}^\dagger 
  +\tfrac14\bar\Lambda^i\gamma_a\Lambda_i\,\mathcal D^0 \big]
  \,\mathcal{H}  \,. 
\end{align}
The class of Lagrangians presented below involves a function
$\mathcal{H}(\phi_\alpha)$ that is homogeneous of zeroth degree in the
holomorphic variables, so that
$\mathcal{D}^\dagger\mathcal{H}(\phi_\alpha)=0$ and $\mathcal{D}^0
\mathcal{H}(\phi_\alpha)=0$. Using the above commutation relations, it
then follows that $\mathcal{D}^\dagger\, \mathcal{D}^n
\mathcal{H}(\phi_\alpha) \propto \mathcal{D}^{n-1}
\mathcal{H}(\phi_\alpha)$ for $n>1$, and vanishes for $n=1$ so that
$\mathcal{D} \mathcal{H}(\phi_\alpha)$ is holomorphic while
$\mathcal{D}^2\mathcal{H}$ is not.

After these definitions we present the bosonic terms of the full class
of superconformal Lagrangians, which are given by the real part of the
following expression, 
\begin{widetext}
\begin{align}
  \label{eq:BosLag}
e^{-1}\mathcal{L} &=
         \mathcal{H}\,\Big[\,
	\tfrac{1}{2}\, {R(M)}^{a b c d}\, R(M)^-_{a b c d}
	+ {R(V)}^{a b}{}^{i}{}_{j} \,R(V)^-_{a b}{}^{j}{}_{i}
	+ \tfrac{1}{8}\, {D}^{i j}{}_{k l} \,{D}^{k l}{}_{i j}
	+ \tfrac{1}{4}\, {E}_{i j} \,D^2 E^{i j}\,
	- 4\, {T}_{a b}{\!}^{i j} {D}^{a}{D}_{c}\,{T}^{cb}{\!}_{i j}
	\nonumber\\[1mm]
	& \qquad\;\;\;
	- {\bar P}^{a} {D}_{a}{D}_{b}{P}^{b}
        + P^2 \bar P^2 + \tfrac{1}{3} (P^a \bar P_a)^2
        - \tfrac{1}{6}\,{P}^{a} {\bar P}_{a} \,{E}_{i j}\,{E}^{i j} 
        - 8\, {P}_{a} \,{\bar P}^{c}\, {T}^{a b}{\!}_{i j} \,{T}_{b c}{\!}^{i j} 
        - \tfrac{1}{16}\, {E}_{i j} \,{E}^{j k} \,{E}_{k l} \,{E}^{li} 
	\nonumber\\[2mm]
        & \qquad\;\;\;
         + \tfrac{1}{48}\, [{E}_{i j}\, {E}^{i j}]^2
	+ {T}^{ab}{\!}_{i j}\, {T}_{a b\, k l}\, {T}^{c d\,i j} \,{T}_{c d}{}^{k l}
	- {T}^{ab}{\!}_{i j}\,{T}_{c d}{\!}^{jk}\, {T}_{a b\, k l}\,{T}^{c d\,l i}
        - \tfrac{1}{2}\, {E}^{i j} \,{T}^{a b\,k l}\, {R(V)}_{a b}{\!}^{m}{}_{i}\, {\varepsilon}_{j k l m}
        \nonumber\\[2mm]
        & \qquad\;\;\;
        	+ \tfrac{1}{2}\, {E}_{i j}\, {T}^{ab}{}_{k l}\,
                {R(V)}_{a b}{\!}^{i}{}_{m} \,{\varepsilon}^{j k l m}  
	- \tfrac{1}{16}\, {E}_{i j} {E}_{k l} \,{T}^{a b}{\!}_{m n}\,
        {T}_{a b \,p q}\,
        {\varepsilon}^{i k m n}\, {\varepsilon}^{j l p q} 
	- \tfrac{1}{16}\, {E}^{i j} {E}^{k l} \,{T}^{a b\,m n} \,{T}_{a b}{}^{p q} \,{\varepsilon}_{i k m n} \,{\varepsilon}_{j l p q} 
	\nonumber\\[2mm]
        & \qquad\;\;\;
	- 2 \,{T}^{a b\,i j}\,\big( {P}_{[a} {D}_{c]}{{T}_{b}{}^{c\,k l}}
        +\tfrac16\,{P}^{c} {D}_{c}{{T}_{a b}{}^{k l}}+\tfrac13\,{T}_{a
          b}{}^{k l} {D}_{c}{{P}^{c}}\big)\,  {\varepsilon}_{i j k l} 
	- 2 \,{T}^{a b}{\!}_{i j} \,\big({\bar P}_{[a} {D}_{c]}{T}_{b}{}^{c}{}_{ k l}
        -\tfrac12\, {\bar P}^{c} {D}_{c}{{T}_{a b \,k l}}\big)\,
        {\varepsilon}^{i j k l}   
        \,	\Big] \nonumber\\[1mm]
        & \;\;\;\,
	+ \mathcal{D}{\mathcal{H}} \,  \Big[\,
	 \tfrac{1}{4}\, {T}_{a b}{}^{i j}\, {T}_{c d}{}^{k l} {R(M)}^{a b c d} \,{\varepsilon}_{i j k l}
	+{E}_{i j} \,{T}^{a b\,i k} \,{R(V)}_{a b}{\!}^{j}{}_{k}
	-\tfrac{1}{8}\, {D}^{i j}{}_{k l} \,\big(\,{T}^{a b\,m n} 
        \,{T}_{a b}{}^{k l}\, {\varepsilon}_{i j m n}
        - \tfrac{1}{2}\, {E}_{i m}\, {E}_{j n}\, {\varepsilon}^{k l m n}\,\big)
	\nonumber\\[1mm]
        & \qquad\;\;\;\;\;\;\;\;\;
        	+{T}^{a b\,i j} \,{T}_{a}{}^{c\,k l} \,{R(V)}_{b c}{\!}^{m}{}_{k} \,{\varepsilon}_{i j l m}
	- \tfrac{1}{24}\, {E}_{i j} \,{E}^{i j} \,{T}^{a b\,k l}\, 
        {T}_{a b}{}^{m n}\, {\varepsilon}_{k l m n}
	- \tfrac{1}{6}\, {E}^{i j} \,{T}_{a b}{}^{k l}\, {T}^{a c\,m n} \,{T}^b{}_{c}{}^{p q} \,{\varepsilon}_{i k l m} \,{\varepsilon}_{j p q n}\,\,
        \Big]	\nonumber\\[1mm] 
        & \;\;\;\; 
        + {\mathcal{D}}^2\mathcal{H}\, \Big[ \tfrac{1}{6}\,
        {E}_{i j} \,{T}_{a b}{}^{i k} \,{T}^{a c \, jl}\,
        {T}^{b}{}_{c}{}^{m n} \,{\varepsilon}_{k l m n} -
        \tfrac{1}{8}\, {E}_{i j} \,{E}_{k l} \,{T}_{a b}{}^{i k}
        \,{T}^{a b \, j l} + \tfrac{1}{384}\, {E}_{i j} \,{E}_{k l}
        \,{E}_{m n} \,{E}_{p q} \,{\varepsilon}^{i k m p}\,
        {\varepsilon}^{j l n q}
	\nonumber\\[1mm]
        & \qquad\;\;\;\;\;\;\;\;\;\;\, 
        + \tfrac{1}{32}\,{T}^{a b \, i
          j}\,{T}^{c d \, p q}\, {T}_{a b}{}^{m n} \,{T}_{c d}{}^{k
          l}\, {\varepsilon}_{i j k l} \,{\varepsilon}_{m n p q}
        -\tfrac{1}{64}\,{T}^{a b \, i j}\,{T}^{c d \, p q}\, {T}_{a
          b}{}^{k l} \,{T}_{c d}{}^{m n} \, {\varepsilon}_{i j k l}
        \,{\varepsilon}_{m n p q} \, \Big]
	\nonumber\\
        &  \;\;\;\;	
	+ 2 \,\mathcal{H} \,e_a{}^\mu f_\mu{}^c\,\eta_{cb} \, \Big[\,
	 {P}^{a} \,{\bar P}^{b}  - P^d\, \bar P_d \, \eta^{a b}\,
	\Big]\,,
\end{align}
\end{widetext}
where $R(M)^-{\!\!\!}_{abcd}$ and $R(V)^-{\!\!\!}_{ab}{}^i{}_j$ denote
anti-selfdual supercovariant curvatures. When suppressing the
fermionic terms and imposing the gauge $b_\mu=0$, they become 
equal to the Weyl tensor and the $\mathrm{SU}(4)$ field strengths.

Let us now turn to the derivation of this result. It makes use of
the fact that any supersymmetric component Lagrangian can be written
as the Hodge dual of a four-form built in terms of the vierbein,
gravitini, and possibly other connections, multiplied by
supercovariant coefficient functions that we will treat as composite
fields. This approach is known as the superform method
\cite{Gates:1997kr,Gates:1997ag}.  Since for $N\!=\!4$ supergravity chiral
superspace does not exist, we aim to construct such a density formula
directly, assuming that only the vierbein and gravitini may appear
explicitly within the four-form. Schematically we will thus consider a
four-form decomposed into five types of forms, namely $\psi^4$,
$e\,\psi^3$, $e^2\, \psi^2$, $e^3\,\psi$ and finally $e^4$. The Weyl
weight of these forms ranges from $w=-2$ for the first one to $w=-4$
for the last one. This last form will be multiplied by a composite
coefficient function with $w=4$ that contains all the purely bosonic
terms of the Lagrangian specified in \eqref{eq:BosLag} (as well as
fermionic terms).

The structure of the Lagrangian is dictated by the transformation of
the lowest-dimensional supercovariant composites.  We thus start by
considering the quartic gravitino forms, which we postulate to be of
the following type,
\begin{align}
  \label{eq:DensityFormula}
  \mathcal{L} &=
        -  \mathrm{i} \,\varepsilon^{\mu\nu\rho\sigma}\,
	\bar\psi_{\mu i}\, \psi_{\nu j} \;\bar\psi_{\rho}{}^k \,\psi_{\sigma}{\!}^l\;
	A^{i j}{\!}_{k l} 
        \nonumber\\ & \quad
	- \tfrac{1}{4} \mathrm{i} \,\varepsilon^{\mu\nu\rho\sigma}\,
	\bar\psi_{\mu i} \,\psi_{\nu j} \;\bar\psi_{\rho k}\, \psi_{\sigma l}\,
	\varepsilon^{k l r s}\,C^{i j}{\!}_{r s}
\nonumber\\ & \quad
	- \tfrac{1}{4} \mathrm{i}\,\varepsilon^{\mu\nu\rho\sigma}\,
	\bar\psi_{\mu}{\!}^i \,\psi_{\nu}{\!}^j\; \bar\psi_{\rho}{\!}^k \,\psi_{\sigma}{\!}^l\,
	\varepsilon_{k l r s}\, \bar C^{r s}{\!}_{i j}
        + \cdots\,,
\end{align}
where the supercovariant composites $A^{ij}{\!}_{kl}$,
$C^{ij}{\!}_{kl}$, and $\bar C^{ij}{\!}_{kl}$ are assumed to be
S-supersymmetric, and are therefore also invariant under conformal
boosts.  All three composites have $w\!=\!2$ and belong to the ${\bf
  20}'$ representation of $\mathrm{SU}(4)$. It turns out that these
are sufficient to generate the full class of superconformal
Lagrangians. It may be possible to include other quartic gravitino
terms in different representations, but this modification will only
give rise to additional total derivatives in the final Lagrangians.

To elucidate the calculation we start by evaluating the
Q-supersymmetry variations that remain proportional to four
gravitini. For this we note that gravitino fields transform under
Q-supersymmetry as
\begin{equation}
  \label{eq:susy-gravitino}
  \delta\psi_\mu{\!}^i =2\, \mathcal{D}_\mu \epsilon^i -\tfrac12
  T_{ab}{\!}^{ij} \gamma^{ab}\gamma_\mu \epsilon_j - 
  \varepsilon^{ijkl}\, \bar\epsilon_j \psi_{\mu k}\,\Lambda_l \,,
\end{equation}
where the derivative $\mathcal{D}_\mu$ is covariant under the bosonic
gauge transformations with the exception of the conformal boosts.  At
the quartic gravitino level we may ignore the second term, but neither
the third nor the first. The reason the first term is relevant is
that, in writing the variation of the action in a manifestly
supercovariant form, we must integrate the derivative by parts and
reconstruct supercovariant quantities.  This then leads to further
gravitino terms in two ways. The first is when the derivative hits
another gravitino, which must be converted into the supercovariant
Q-supersymmetry curvature by adding appropriate terms,
\begin{align}
  \label{eq:D-psi}
  \mathcal D_{[\mu} \psi_{\nu]}{}^i &=
	\tfrac{1}{2} R(Q)_{\mu\nu}{}^i
	- \tfrac{1}{4} \varepsilon^{i j k l} \bar \psi_{[\mu j} \psi_{\nu] k} \Lambda_{l}
	+ \cdots~.
\end{align}
The second way is when the derivative hits a supercovariant composite, such as
$C^{ij}{\!}_{kl}$, which we rewrite as
\begin{equation}
  \label{eq:cov-der-Xi}
  \mathcal{D}_\mu C^{ij}{\!}_{kl} = D_\mu C^{ij}{\!}_{kl} + \tfrac12
  \big[\bar\psi_\mu{\!}^m \,\Xi^{ij}{\!}_{kl,m} +
  \bar\psi_{\mu m} \,\Xi^{ij,m}{\!}_{kl} \big]\, .
\end{equation}
Here we write the Q-supersymmetry transformations of the
scalar composites as
\begin{align}
  \delta C^{ij}{\!}_{kl} &= \bar\epsilon^m \,\Xi^{ij}{\!}_{kl,m} +
  \bar\epsilon_m\,\Xi^{ij,m}{\!}_{kl} \,, \nonumber \\
  \delta A^{ij}{\!}_{kl} &= \bar\epsilon^m \,\Omega^{ij}{\!}_{kl,m} +
  \bar\epsilon_m\,\Omega^{ij,m}{\!}_{kl} \,,
\label{eq:deltaCA}
\end{align}
with fermionic composites $\Xi$ and $\Omega$. Naturally the
transformations \eqref{eq:deltaCA} also induce variations of
\eqref{eq:DensityFormula} proportional to $\psi^4$ times $\Xi$ and
$\Omega$. Finally there is yet another way to generate variations
quartic in gravitini originating from a four-form of the type
$e\,\psi^3$, induced by the transformation $\delta e_\mu{\!}^a=
\bar\epsilon^k \gamma^a \psi_{\mu k} +\bar\epsilon_k \gamma^a
\psi_{\mu}{\!}^k$. Note that connections other than the gravitini will
also be generated, for example from \eqref{eq:D-psi}, but those turn
out to cancel at the end.

Collecting all the resulting variations proportional to the various
possible quartic gravitini four-forms and requiring them to vanish
imposes the following constraints on the traceless parts of the
fermionic composites in \eqref{eq:deltaCA},
\begin{align} 
  [\Xi^{i j}{\!}_{kl,m}]_{\overline{\bf 60}} &= [2
  \Lambda_m A^{i j}{}_{kl}]_{\overline{\bf 60}}~, \quad [\Xi^{ij,
    m}{\!}_{kl}]_{{\bf 60}} = 0~, \qquad
  \nonumber \\
  [\Omega^{i j}{}_{kl,m}]_{\overline{\bf{60}}} &= [\Lambda_m \bar C^{i
    j}{}_{kl}]_{\overline{\bf {60}}}~,
\label{eq:Bigconstraint}
\end{align}
along with their complex conjugates, where $[\bullet]_{\bf r}$ denotes
projection onto the $\mathrm{SU}(4)$ representation $\bf r$.  The
remaining terms in the fermionic composites $\Xi$ and $\Omega$ lie in
the $\bf{20}$ and $\overline{\bf{20}}$ representations and must be proportional to the fermionic composites multiplying the $e\,\psi^3$
four-forms. We refrain from giving explicit formulae as the general
pattern should be clear. Note that so far we have only made use of the
transformations of the vierbein and the gravitini.

This procedure must be continued by considering the remaining
Q-supersymmetry variations proportional to the four-forms $e\,\psi^3$,
$e^2\,\psi^2$, etc., to determine the relations between all the
supercovariant composites in the Lagrangian and their transformation
rules. This calculation will make use of the Q-supersymmetry
transformations of almost all the Weyl multiplet fields specified in
\cite{Bergshoeff:1980is,Ciceri:2015qpa}. The fact that no
inconsistencies arise at this level is a first indication that our
original assumptions regarding the $\psi^4$ four-forms are correct.

Finally, we must check that the derived transformation rules of the
composites satisfy the same off-shell superconformal algebra as the
Weyl multiplet. This is a straightforward but technically involved
calculation for which we made extensive use of the computer algebra
package {\it Cadabra} \cite{Peeters:2007wn, Peeters:2006kp}. In this
process one also identifies the missing S-supersymmetry variations of
the composites, which so far were only specified for $A^{ij}{\!}_{kl}$
and $C^{ij}{\!}_{kl}$.  This then provides a complete density formula
built upon \eqref{eq:DensityFormula} and \eqref{eq:Bigconstraint},
that is invariant under all local superconformal symmetries. We
emphasize that this density formula makes no assumptions about the
specific dependence of the supercovariant composites on the Weyl
multiplet fields.

At this point we have fully confirmed all the assumptions that are at
the basis for the above calculations. To obtain the final results we
express the composites $A^{ij}{\!}_{kl}$ and $C^{ij}{\!}_{kl}$ in
terms of the supercovariant fields of the $N\!=\!4$ Weyl multiplet;
they then yield corresponding expressions for all the composites
through their transformation rules. As it turns out there exist only
four scalar S-supersymmetric expressions in the ${\bf20}'$
representation,
\begin{align}
  \label{eq:decmp-A-C}
X_{(1)}{\!}^{ij}{\!}_{kl} &= D^{ij}{\!}_{kl}~, \nonumber\\
X_{(2)}{\!}^{ij}{\!}_{kl} &= 
	\tfrac{1}{2} T_{ab}{\!}^{ij} T^{ab\,mn} \varepsilon_{klmn}
	- \tfrac{1}{4} \varepsilon^{ijmn} E_{mk} E_{nl} \nonumber\\
	& \quad
	+ 2 \bar\Lambda_{[k} \chi^{ij}{\!}_{l]}
	- \text{traces}~,\nonumber\\
X_{(3)}{\!}^{ij}{\!}_{kl} &= 
	- \tfrac{1}{4} T_{ab}{\!}^{ij} \bar\Lambda_k \gamma^{ab} \Lambda_l 
	+ \tfrac{1}{4} \varepsilon^{ijmn} E_{m [k} \bar\Lambda_{l]} \Lambda_n
	- \text{traces}~,\nonumber\\
X_{(4)}{\!}^{ij}{\!}_{kl} &=
	-\tfrac{1}{24} \varepsilon^{ijmn} \bar\Lambda_k \Lambda_m
        \bar\Lambda_n \Lambda_l~, 
\end{align}
each of which is homogeneous in the Weyl multiplet fields. The first
one is pseudo-real and the others are complex.  The composites
$A^{ij}{\!}_{kl}$ and $C^{ij}{\!}_{kl}$ must then be written as linear
combinations of $X_{(n)}{\!}^{i j}{\!}_{kl}$ and $\bar X_{(n)}{\!}^{i
  j}{\!}_{kl}$ multiplied by apriori arbitrary functions of the coset
scalars with the appropriate $\mathrm{U}(1)$ weights.
From these expressions one determines the corresponding fermionic
composites via \eqref{eq:deltaCA}, and subsequently imposes the
constraints \eqref{eq:Bigconstraint}.  This then leads to two linearly
independent solutions for $A^{ij}{\!}_{kl}$ and $C^{ij}{\!}_{kl}$. 

One solution turns out to correspond to a Lagrangian that is a total
derivative. The other one depends on an arbitrary holomorphic function
$\mathcal{H}(\phi_\alpha)$ that is homogeneous of zeroth degree and an
associated potential $\mathcal{K}(\phi_\alpha, \phi^\beta)$, which
obeys $\mathcal D \mathcal{D}^\dagger \mathcal{K} = \mathcal{H}$. The
function $\mathcal H$ is uniquely determined as it appears in
$C^{ij}{\!}_{kl}$ as a distinctive term equal to $\tfrac12\mathrm{i}\,
X_{(2)}{\!}^{ij}{\!}_{kl} \,\mathcal{H}$. The holomorphicity of
$\mathcal{H}$ can then be seen as a direct consequence of the
constraints \eqref{eq:Bigconstraint}, which also determine how
derivatives of $\mathcal H$ and $\mathcal{K}$ appear within
$C^{ij}{\!}_{kl}$ and the other composites. At the end the potential
$\mathcal{K}$ is removed by splitting off a total derivative. In
deriving the result \eqref{eq:BosLag} we have introduced additional
total derivative terms in order to bring the formula into a concise
form at the cost of generating terms that explicitly depend on the
conformal boost gauge field $f_\mu{\!}^a$.

In this Letter we have described the construction of a class of
Lagrangians of $N\!=\!4$ conformal supergravity encoded in an
arbitrary holomorphic function that is homogeneous of zeroth
degree. Up to total derivative terms these are the only invariant
Lagrangians that exist. We will return to this point in a forthcoming
paper where we will also present the details of this calculation,
including the full density formula and all relevant supersymmetry
transformations. In this construction many consistency checks were
carried out. Perhaps the most stringent one is the comparison of the
bosonic Lagrangian \eqref{eq:BosLag} with the case that the
holomorphic function equals a constant, to the result of
\cite{Ciceri:2015qpa}. Indeed both expressions agree up to a total
derivative.

The Lagrangians derived above can be directly incorporated into
Poincar\'e supergravity as a four-derivative coupling following the
same construction carried out originally in \cite{deRoo:1984zyh}, by
including the superconformal Lagrangian of this Letter before
proceeding to the standard gauge choices. The $\mathrm{SU}(1,1)$
symmetry will then become entangled with an electric-magnetic duality
transformation in the vector-multiplet sector. Alternatively this
Lagrangian can be obtained in one step by applying the superform
method directly based on an extended field configuration consisting of
at least six (on-shell) vector supermultiplets and the Weyl
supermultiplet, where one must bear in mind that the supersymmetry
algebra for this field representation will no longer close off shell.
It is then possible to compare the corresponding expressions with the
$R^2$-couplings for $N\!=\!4$ Poincar\'e supergravity of
\cite{Bossard:2012xs,Bossard:2013rza}, which may also depend
non-trivially on the coset fields.

In the context of Poincar\'e supergravity the higher-derivative
couplings are primarily studied as potential counter\-terms that could
render the theory finite. At this moment there is agreement that this
theory is not finite at the four-loop level \cite{Bern:2013uka}. The
presence of a $\mathrm{U}(1)$ anomaly and the non-trivial dependence
on the coset fields plays an important role in this discussion, as was
extensively discussed in \cite{Carrasco:2013ypa}.

As we mentioned earlier, another possible application of the result of
this Letter concerns the calculation of the corrections to $N\!=\!4$
supersymmetric black hole entropy that are known to originate from
precisely this class of Lagrangians. In principle this can be done by
generalizing the analysis of \cite{LopesCardoso:1999fsj}, which can be
utilized for localization along the lines of
\cite{Dabholkar:2010uh,Murthy:2015yfa}. Both these approaches have
only been applied so far to $N\!=\!2$ supersymmetric truncations. It
should be very interesting to understand these results in the context
of a manifestly $N\!=\!4$ supersymmetric formulation.

\begin{acknowledgments}
  We thank Kasper Peeters for frequent correspondence regarding the
  use of {\it Cadabra}, and Zvi Bern, Guillaume Bossard, Renata Kallosh
  and Kellogg Stelle for informative discussions on anomalies and
  divergences of $N\!=\!4$ supergravity. B.S. thanks Nikhef Amsterdam
  for hospitality extended to him during the course of this work. We
  thank the 2015 Simons Summer Workshop where this work was initiated,
  and the 2016 ``Supergravity: what next?'' workshop at the Galileo
  Galilei Institute for Theoretical Physics where it was completed,
  and the INFN for partial support.  This work is also partially
  supported by the ERC Advanced Grant no. 246974, {\it
    ``Supersymmetry: a window to non-perturbative physics''} and by
  the Marie Curie fellowship PIIF-GA-2012-627976.
\end{acknowledgments}

\providecommand{\href}[2]{#2}\begingroup\raggedright\endgroup

\end{document}